\documentclass[review]{elsarticle}

\usepackage{amsmath}
\usepackage{graphicx}
\usepackage{url}
\usepackage{epstopdf}

\journal{Journal of \LaTeX\ Templates}

\begin{document}

\begin{frontmatter}

\title{Novel method for hit-position reconstruction using voltage signals in plastic scintillators and its application to Positron Emission Tomography}

\author[SWIERK]{L.~Raczy\'nski}
\author[WFAIS]{P.~Moskal}
\author[SWIERK]{P.~Kowalski}
\author[SWIERK]{W.~Wi\'slicki}
\author[WFAIS]{T.~Bednarski}
\author[WFAIS]{P.~Bia\l as}
\author[WFAIS]{E.~Czerwi\'nski}
\author[WFAIS,PAN]{\L .~Kap\l on}
\author[WCHUJ]{A.~Kochanowski}
\author[WFAIS]{G.~Korcyl} 
\author[WFAIS]{J.~Kowal}
\author[WFAIS]{T.~Kozik}
\author[WFAIS]{W.~Krzemie\'n}
\author[WFAIS]{E.~Kubicz}
\author[WCHUJ]{M.~Molenda}
\author[WFAIS]{I.~Moskal}
\author[WFAIS]{Sz.~Nied\'zwiecki}
\author[WFAIS]{M.~Pa\l ka}
\author[WFAIS]{M.~Pawlik-Nied\'zwiecka}
\author[WFAIS]{Z.~Rudy}
\author[WFAIS]{P.~Salabura}
\author[WFAIS]{N.G.~Sharma}
\author[WFAIS]{M.~Silarski}
\author[WFAIS]{A.~S\l omski} 
\author[WFAIS]{J.~Smyrski}
\author[WFAIS]{A.~Strzelecki}
\author[WFAIS,PAN]{A.~Wieczorek}
\author[WFAIS]{M.~Zieli\'nski}
\author[WFAIS]{N.~Zo\'n}

\address[SWIERK]{\'Swierk Computing Centre, National Centre for Nuclear Research, 05-400 Otwock-\'Swierk, Poland}
\address[WFAIS]{Faculty of Physics, Astronomy and Applied Computer Science,
 Jagiellonian University, 30-059 Cracow, Poland}
\address[PAN]{Institute of Metallurgy and Materials Science of Polish Academy of Sciences, Cracow, Poland.}
\address[WCHUJ]{Faculty of Chemistry, Jagiellonian University, 30-060 Cracow, Poland}

\begin{abstract}
Currently inorganic scintillator detectors are used in all commercial Time of Flight Positron Emission Tomograph 
(TOF-PET) devices. The J-PET collaboration investigates a possibility of construction of a PET scanner from plastic scintillators which would allow for single bed imaging of the whole human body. This paper describes a novel method of hit-position reconstruction based on sampled signals and an example of an application of the method for a single module with a 30~cm long plastic strip, read out on both ends by Hamamatsu R4998 photomultipliers. The sampling scheme to generate a vector with samples of a PET event waveform with respect to four user-defined amplitudes is introduced. The experimental setup provides irradiation of a chosen position in the plastic scintillator strip with an annihilation gamma quanta of energy 511~keV. The statistical test for a multivariate normal (MVN) distribution of measured vectors at a given position is developed, and it is shown that signals sampled at four thresholds in a voltage domain are approximately normally distributed variables. With the presented method of a vector analysis made out of waveform samples acquired with four thresholds, we obtain a spatial resolution of about 1~cm and a timing resolution of about 80~ps ($\sigma$).

\end{abstract}

\begin{keyword}
\texttt{Positron Emission Tomography} \sep \texttt{plastic scintillators} \sep \texttt{statistical analysis}
\end{keyword}

\end{frontmatter}

\section{Introduction}
Positron Emission Tomography (PET) \cite{Bailey2005,Humm2003} represents one of the most prominent perspective techniques of non-invasive imaging in medicine. The first demonstration of recording data in positron detection was taken in early 1950s \cite{Brownell1953}, only two years after the first medical application of the positron was reported \cite{Sweet1951}. In 1973 J.~Roberston and his co-workers built the first ring PET tomograph, which consisted of 32 detectors  \cite{Robertson1973}. This scanner has become the prototype of the current shape of PET.

Since the early detection of small lesions and  monitoring of the functionality of organs are critical for prophylaxis and efficient treatments, notable efforts are nowadays devoted to improve the resolution of reconstructed images. It was realized that the measurement of the difference of arrival times, or times of flight (TOF), of two gamma rays provides substantial progress in this domain \cite{Karp2008,Kardmas2009,Townsend2004,Moses1999,Moses2003}. The new class of instruments, called TOF-PET, better localize the emission source along a straight line of gamma coincidence, called the Line of Response (LOR). The LORs are basic components of image reconstruction algorithms.

Currently all commercial PET devices use inorganic scintillator materials, usually LSO or LYSO crystals, as radiation detectors. These are characterized by relatively long rise- and decay times, of the order of tens of nanoseconds. Time resolution in PET examinations is crucial and one observes persistent strive for improvement
~\cite{Conti2009,Schaart2010,MoszynskiSzczesniak2011,Szczesniak2010,AmpSampl,AX-PET}.

In recent articles ~\cite{NovelDetectorSystems,StripPETconcept,TOFPETDetector,JPET-Genewa}, an utterly new concept of TOF-PET scanner was introduced. It incorporates plastic scintillators with good resolving time and the TOF method. Disadvantages due to the low detection efficiency and negligible photoelectric effect in organic scintillators can be compensated by large acceptance and good time resolution~\cite{TOFPETDetector}. In addition, the method allows for  drastic reduction of the production cost of PET scanners and is promising for the construction of the single bed whole body PET scanner. A single detection unit of the newly proposed TOF-PET detector~\cite{StripPETconcept} is built out of a long strip of scintillator, read out on both sides by photomultipliers. Such a solution enables the reconstruction of coordinates of the gamma quantum interaction along the scintillator strip by measuring signals at its two ends. A similar solution for crystal scintillators has been recently developed by the AX-PET collaboration~\cite{AX-PET}. The 10~cm long LYSO crystals were coupled to digital Silicon Photomultipliers (dSiPM), and a very good coincidence time resolution of about 200~ps (FWHM) was achieved. In plastic scintillators, the 511~keV quanta from electron-positron annihilation produces signals burdened with large fluctuations of the number of photoelectrons. Therefore a usage of the typical techniques for time measurement, based on the application of a single-level leading-edge or constant-fraction discriminators, is not sufficient. Therefore a multithreshold sampling method to generate samples of a PET event waveform with respect to four user-defined amplitudes was proposed. A similar idea may be found in~\cite{AmpSampl}, where a coincidence timing resolution of about 340~ps was obtained for LSO crystals. In~\cite{AmpSampl} the four sample points that comprise the rising edge were fit to a line, and the intersection of the fitted line with the zero voltage level defined the event time of the pulse. 

In this article we propose a new method for reconstruction of the gamma quantum hit position. The method is based on the statistical model of signals probed in the voltage domain. An electronic system for probing these signals is under 
development~\cite{palka}. In the following we describe a new concept of reconstruction of the gamma quantum hit position. The description includes an explanation of the methods used for the test of the normality of data, determination of the effective number of degrees of freedom, as well as explanation of the selection criteria applied to the experimental sample. Then we describe an experimental setup used for signal registration and present results of reconstruction of the hit positions of gamma quanta in 30~cm long plastic scintillator strip, read out on both sides by the Hamamtsu photomultipliers R4998. Signals from the photomultipliers were sampled in 50~ps steps using the Lecroy Signal Data Analyzer 6000A.

\section{Description of the reconstruction method}

Light pulses produced in the strip propagate to its edges where they are converted into electric signals. These are sampled in the amplitude domain using a predefined number of voltage levels. The measurement gives a vector of $N$ values of times at which the signal crosses the reference voltages. This vector carries information about the shape of the signals and their times of arrivals to the edges of the scintillator. These shapes depend on the hit position and can be used for its reconstruction. 

The method of hit-position reconstruction consists of two steps. First, the scintillator's volume is discretized and for each bin a high statistics set of reference signals is created. In the example discussed later, each set contains approximately 5~000 signals generated by irradiation with gamma quanta at a fixed scintillator position. The objective of the second part of the procedure is to qualify the new measurement to one of the given sets of signals and hence determine the hit position.

Consider $L$ data sets $S_{i}$ ($i = 1,.., L$). Each $S_{i}$ is a $M_i x N_i$ matrix of vectors representing signals gathered for the $i^{\text{th}}$ position; $M_i$ is the number of the collected signals and $N_i$ stands for vector's dimension equal to the number of samples per signal. In practice all signals have the same dimension and $N_{i} =N$ for all $i$. The $j^{\text{th}}$ signal in the $i^{\text{th}}$ data set corresponds to the $j^{\text{th}}$ row of the matrix $S_{i}$ and is denoted by the vector $s_{j}^{(i)}$. If the measured coordinates of vectors in all $L$ data sets are normally distributed then the mean value $m_i$ and covariance matrix $C_i$ of the data set $S_{i}$ describe it completely.

Assuming their normality, the proposed reconstruction procedure qualifies a new measurement, represented by vector $u$, to one of the data sets $S_{i}$ by using only information about $m_i$ and $C_i$. In the first step of the reconstruction, the Mahalanobis distances $d^{(i)}$ between $u$ and $m_i$ are calculated:
\begin{equation}
	d^{(i)} = (u - m_{i}) \cdot C_{i}^{-1} \cdot (u - m_{i})^{\text{T}} ~~~~~~~~  i = 1, 2, ..., L.
        \label{klasyf}
\end{equation}
Next, the measured signal $u$ is qualified to the data set $i^{*}$ with the smallest distance $d^{(i)}$:
\begin{equation}
	i^{*} = \arg(\min( d^{(i)} )).
        \label{i_hat}
\end{equation}

\subsection{Test of the normality of data}

There are number of existing procedures for testing whether multivariate vectors from a given dataset have a multivariate normal (MVN) distribution. For example Mardia~\cite{Mardia} proposes multivariate measures of skewness and kurtosis, which are special cases of MVN moment restrictions. In addition, multivariate generalization of the well known Kolmogorov-Smirnov~\cite{Lillie} and Shapiro-Wilks~\cite{Fatto,Villa} tests have been established. There are also available tests based on a $\chi^2$ quantile-quantile (q-q) plot of the observations squared Mahalanobis distances. 

In this work, we propose an alternative procedure for testing a MVN distribution as an extension of statistical test based on q-q approach~\cite{qqplot}. In order to verify normality of the data set in $S_i$, the observations squared Mahalanobis distances for $M_i$ vectors from $S_i$ data set are calculated:
\begin{equation}
	d_{j}^{(i)} = (s_{j}^{(i)} - m_{i}) \cdot C_{i}^{-1} \cdot (s_{j}^{(i)} - m_{i})^{\text{T}} ~~~~~~~~  j = 1, 2, ..., M_{i}.
        \label{Mahal_dist}
\end{equation}
where $m_i$, and $C_i$ are estimated based on the data set $S_i$. In~\cite{Mulnor} authors assumed that the evaluated distances in Eq.~\ref{Mahal_dist} have a $\chi^2$ distribution with $N$ degrees of freedom. In the following we will show that this is not necessarily the case, and the number of effective degrees of freedom may be smaller due to signal correlation. The discussion about the effective number of degrees of freedom (denoted hereafter as $V$) will be given in next Section. 

We provide a statistical test for data set $S_i$ by comparing the distribution of $d_{j}^{(i)}$ defined in Eq.~\ref{Mahal_dist} with the theoretical $\chi^2$ distribution with $V$ degrees of freedom. The normalization of theoretical histogram is provided to ensure that sum of counts in both histograms is the same and equal to $M_i$ (see Eq.~\ref{Mahal_dist}). We apply uneven bin size, in order to store in each bin of the theoretical $\chi^2$ histogram a constant number of counts $F_T$. This simplifies a control of the assumption about the normal distribution of number of counts in each bin. In the calculations, we have selected $F_T$ = 30, and therefore the Poisson distribution may be approximated accurately by the normal distribution. Hence, we compare the two histograms via statistical test $R$ defined as follows:
\begin{equation}
	R_{i}(V) = \sum_{k=1}^{K_{i}}\frac{ F_{k}^{(i)} - F_{T} }{ F_{T} },
	\label{test_R}
\end{equation}
where $F_{k}^{(i)}$ value is the number of counts in the $k^{\text{th}}$ bin in the experimental histogram from the $i^{\text{th}}$ data set, and $K_i = M_i /F_T$ (the number of bins in the histograms from the $i^{\text{th}}$ data set). The bin sizes were calculated from the theoretical $\chi^2$ with $V$ degrees of freedom. The test statistic $R_i$ is a chi-squared random variable with mean $K_i$ and standard deviation $ \sqrt{2 K_{i}} $; owing to well known concentration inequalities, the probability that $R_i$ exceeds its mean plus three standard deviations is small. In the following we will find the parameter $\lambda$ that fulfills the equation: 
\begin{equation}
	R_{i}(V) = K_{i} + \lambda \sqrt{2 K_{i}},
	\label{lambda}
\end{equation}
and we state that the null hypothesis that the experimental histogram has a $\chi^2$ distribution with $V$ degrees of 
freedom is true if $\lambda < 3$. 

We wish to make one comment about the practical application of the test proposed in this Section. The number of collected signals ($M_i$) and hence the number of bins ($K_i$) in Eq.~\ref{test_R}, should be large enough to describe properly the smooth function $\chi^2$ with $V$ degrees of freedom. We will not provide an evaluation of a minimal number of bins, for arbitrary chosen $F_T$ = 30, but we suggest to use the test in the case of large data sets with $M_{i} > $ 1000.

\subsection{Number of effective degrees of freedom}

Components of the signal vector are mutually correlated in a complicated manner so the effective $V$ has to be determined empirically. Its upper bound $V_{\text{max}}$ is equal to the number of independent variables $N$. In order to determine the minimal $V_{\text{min}}$, the Principal Component Analysis (PCA)~\cite{PCA} of data set $S_i$ is performed. Before full PCA examinations, the column means of the data set $S_i$ are subtracted, in order to standardize distributions of the vectors' components. The data set with 0 mean value will be depicted with $S_i^0$. We define the orthonormal matrix $ W_i \in R^{NxN} $ that maps the vectors from data set $S_i^0$ into a new space,
\begin{equation}
 	\hat S_i = S_i^0 \cdot W_i, 
\end{equation}
in such a way that the projection with successive basis vectors inherits the greatest possible variance in
data set $S_i^0$. The covariance matrix of data set $\hat S_i$ will be denoted with $\hat C_i$ and is given as 
$ \hat C_i = E(\hat S_i^{\text{T}} \cdot \hat S_i).$ It is diagonal, with values sorted in non-increasing order. We define the Total Variance ($TV$) parameter as a normalized sum of $k$ variances on the diagonal of $\hat C_i$,
\begin{equation}
	TV_{k} = O_k \cdot \hat C_i \cdot O_k^T \cdot (O_N \cdot \hat C_i \cdot O_N^T)^{-1},
	\label{TV}
\end{equation}
where $O_k$ is the $N$-dimensional row vector with ones at positions from 1 to $k$, and zeros from $k$+1 to $N$. According to this definition, $O_N$ is a vector with all $N$ values equal to one. The $TV$ is a non-decreasing function. We assume that at least $TV > 0.95$ is necessary to describe data set $S_i$ properly. The minimal number of variables $V_{\text{min}}$ is equal to the smallest $k$ for which $TV_k > 0.95$. 

After the determination of $V_{\text{min}}$, calculations of statistics $R$ are repeated for different $V$ in the range from $V_{\text{min}}$ to $V_{\text{max}}$. The theoretical $\chi^2$ distribution with $V$ degrees of freedom for which the smallest statistics $R$ 
(Eq.~\ref{test_R} ) and hence smaller parameter $\lambda$ (from Eq.~\ref{lambda}) was calculated, is selected. The experimental distribution is said to be a MVN distribution with $V$ degrees of freedom, if $\lambda$ is smaller than 3.

\subsection{Method for data cleaning}

If the data are normally distributed, the statistical significance of assignment of the measurement $u$ to data set $S_i$ can
 be provided. The Mahalanobis distance $d$ in Eq.~\ref{Mahal_dist}, from the $\chi^2$ distribution with $V$ of degrees of freedom, can be interpreted using $p$-values~\cite{person}. The hypothesis that $u$ may be assigned to $S_i$ is rejected when the 
 $p$-value is below the predetermined significance level (e.g. 0.01), indicating that the signal $u$ is very unlikely under this hypothesis. Equivalently, a threshold on the Mahalanobis distance $d$, depicted with $d_{\text{max}}$, ensuring a minimal expected 
 $p$-value, can be provided. Finally, the measurement $u$ is qualified to the data set $i$ if, and only if, the distance $d$ to $S_i$ is smaller than the predefined threshold $d_{\text{max}}$. In practice, the $d$- or $p$-value criterion may be used for the rejection of background events due to e.g. signal distortion by gamma quantum rescattering~\cite{Szym}.

\section{Experimental Setup}

\begin{figure}[h!]
	\centerline{\includegraphics[width=0.7\textwidth]{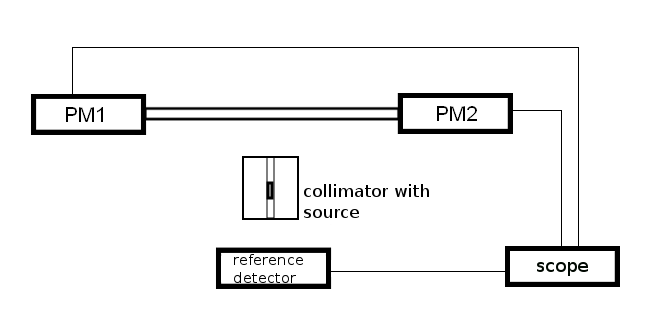}}
\caption{
Scheme of the experimental setup. 
\label{Exper_setup}
}
\end{figure}

\begin{figure}[h!]
	\centerline{\includegraphics[width=1.0\textwidth]{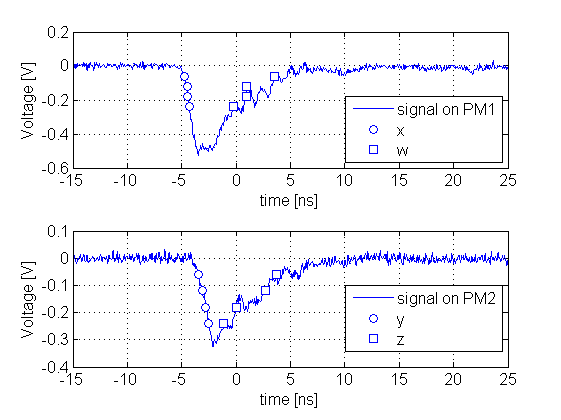}}
\caption{
Example of signals registered at two photomultipliers. Squares and circles denote points at the signal for the voltage values of 60, 120, 180 and 240 mV. The meaning of variables $x$, $w$, $y$, $z$ is described in the text.
\label{Signal_examp}
}
\end{figure}

The method described in the previous sections was tested on the example of reconstruction of hit-position in a single module of the J-PET detector~\cite{StripPETconcept}. The measurement was performed with a single module consisting of the 30~cm plastic scintillator strip EJ-230~\cite{EJ} with the rectangular profile ~1.9~cm~x~0.5~cm. The strip was connected on two sides, via optical gel, to the Hamamatsu photomultipliers R4998, denoted as PM1(2) in Fig.~\ref{Exper_setup}. A series of measurements was performed using collimated gamma quanta from $^{22}$Na source placed between the scintillator strip and reference detector. A collimator was located on a dedicated mechanical platform allowing one to shift it along the line parallel to the scintillator strip with a submillimeter precision. 
The $^{22}$Na source was moved from the first to the second end in steps of 6~mm. At each position, about 5~000 pairs of signals from PM1 and PM2 were registered in coincidence. These signals were sampled using the Signal Data Analyzer 6000A with a probing interval of 50~ps. As a trigger, a coincidence between signals from PM1 and a reference detector was required. Such trigger conditions, together with the 1.5~mm slit in a 20~cm long lead collimator, enabled us to select annihilation quanta. The background of gamma quanta from deexcitation of $^{22}$Ne is less than $0.1\%$. The size of the spatial profile of such triggered annihilation quanta was determined to be about 2~mm (FWHM). Examples of two signals registered at PM1 and PM2 are shown in Fig.~\ref{Signal_examp}. The upper (lower) panel of Fig.~\ref{Signal_examp} shows a signal registered in PM1(2) for the case when the scintillator was irradiated at 
7 and 23~cm to PM1 and PM2,~respectively.

\begin{figure}[h!]
\begin{tabular}{c c}
	\includegraphics[width=0.49\textwidth]{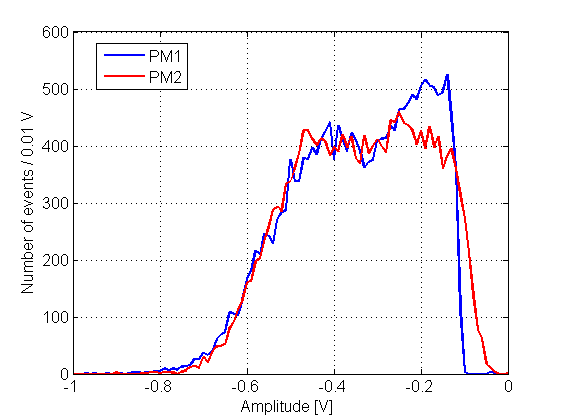} & \includegraphics[width=0.49\textwidth]{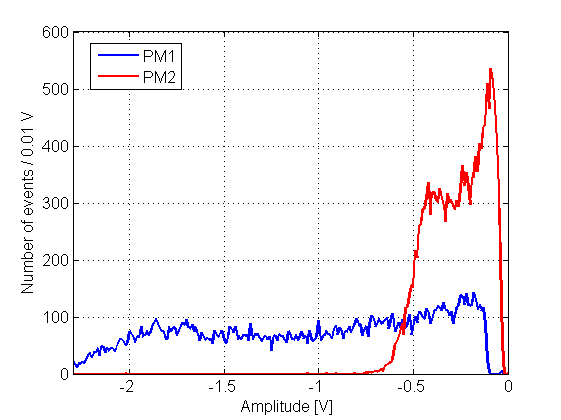}\\
\end{tabular}
\caption{
Distributions of the amplitude signals gathered at central (left panel) and left end (right panel) position of the strip. A sharp edge of the spectrum for the PM1 is due to the triggering conditions, as described in the text.
\label{Ampl_distrib}
}
\end{figure}

In the first step of the analysis the distributions of signal amplitudes were investigated. Experimental results based on the signals registered at the center and end positions of the scintillator strip, are presented in Fig.~\ref{Ampl_distrib} on the left and right panel, respectively, where signals from PM1(2) are shown in blue (red). They reflect the energy distribution of electrons scattered by gamma quanta via the Compton effect. Due to the low atomic number of the elements in plastic (carbon and hydrogen), the maximum from the photoelectric effect is not seen. As expected, the amplitude distributions for two PMs are very similar for central irradiation, and differ significantly for exposition closer to one of the PMs.

\section{Application of the hit-position reconstruction method to the experimental data}

Though in principle all points from the sampled signal could be used for the reconstruction, in practice, in case of hundreds of detection modules, the front-end electronics are able to perform at only a few samples per signal. Therefore in the following, the signals are probed at four fixed-voltage levels, providing eight time values for each signal from PM1 or PM2 - enough to estimate the resolution of hit positions.

\subsection{Choice of variables}

Based on the measurements of fully sampled signals, we simulate a four-level measurement with sampling in the voltage domain at 60, 120, 180 and 240~mV (see Fig.~\ref{Signal_examp}). Sampling times from PM1(2) at a given level are denoted with $x$ ($y$) and $w$ ($z$), for the rising and falling slope, respectively (Fig.~\ref{Signal_examp}) – in total 16 values for two signals, where only differences of times are physically meaningful. The effective number of registered variables is thus smaller by one and equal to $N = 15$. Hence, each data set contains signals registered at a specified position, where single measurement is represented by a 15-dimensional real number vector.   

From the point of view of further reconstruction procedure, all linear combinations of time values are equivalent to each other. We use the following ones:
\begin{equation}
	x_i - y_i ~~~~~~~~~~~~~~ i = 1,..,4,
	\label{x_y}
\end{equation}
\begin{equation}
	x_i - x_{i+1} ~~~~~~~~~~ i = 1,..,3,
	\label{x_x}
\end{equation}
\begin{equation}
	x_i - w_i ~~~~~~~~~~~~~~ i = 1,..,4,
	\label{x_w}
\end{equation}
\begin{equation}
	y_i - z_i ~~~~~~~~~~~~~~ i = 1,..,4,
	\label{y_z}
\end{equation}
They correspond to: time difference of signals from two PMTs at a given voltage levels (Eq.~\ref{x_y}), time differences of a signal from PM1 at adjacent levels (Eq.~\ref{x_x}), and width of the signal on PM1 and PM2 at given levels (Eqs.~\ref{x_w} and \ref{y_z}, respectively). According to this choice, the $j^{\text{th}}$ measurement in the $i^{\text{th}}$ data set may be represented as: $s_j^{(i)} = [x_1-y_1, x_2-y_2, x_3-y_3, x_4-y_4, x_1-x_2, x_2-x_3, x_3-x_4, x_1-w_1, x_2-w_2, x_3-w_3, x_4-w_4, y_1 -z_1, y_2-z_2, y_3-z_3, y_4-z_4]$.
	
The signals  were registered in 50~ps steps (blue curves in Fig.~\ref{Signal_examp}). In order to evaluate the time value at given thresholds, interpolation must be applied. Due to the very high sampling rate, results obtained with different interpolation methods (linear, spline \cite{Forsyt}) were found to be very close to each other. We used the linear interpolation in order to minimize computational cost.

\subsection{Number of effective degrees of freedom}

According to the procedure described in Sec. 2, the PCA is performed and subsequently the $TV$ is determined as a function of the number of variables. The result is shown in Fig.~\ref{TV}.

\begin{figure}[h!]
	\centerline{\includegraphics[width=0.7\textwidth]{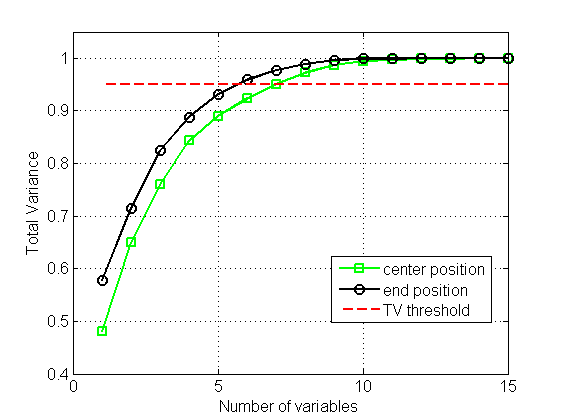}}
\caption{
Examples of the Total Variance determined as a function of the assumed number of independent variables. Green points indicate the result for central irradiation and black stand for the marginal one. The horizontal line indicates the criterion for the determination of the minimum $V$ and curves are to guide one's eye.
\label{TV}
}
\end{figure}

Two curves representing signals registered in two most peripheral places are very similar. The minimal $V$ of the $\chi^2$ statistics ($V_{\text{min}}$) is the argument of the $TV$ function crossing the threshold line, marked in red in Fig.~\ref{TV}. From Fig.~\ref{TV}, the $V_{\text{min}}$ value of 6 and 8 may be read for the data samples from the end and center positions, respectively. To make sure that the condition $TV > 0.95$ is fulfilled in all cases, the $V_{\text{min}}$ value equal to 8 is selected for further studies.

\subsection{Validation of the normality of the collected data sample}

The hypothesis of normality was tested for numbers of degrees of freedom ranging from  
$V_{\text{min}} = 8$ to 15. The comparison of the experimental distribution with the theoretical one was performed based on the statistical test $R$ defined in Eq.~\ref{test_R}. 

\begin{figure}[h!]
	\centerline{\includegraphics[width=0.7\textwidth]{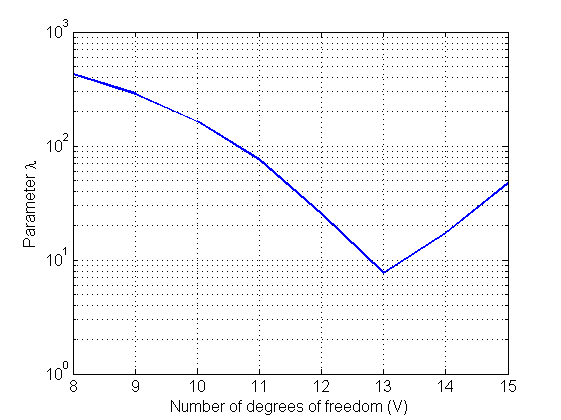}}
\caption{
Parameter $\lambda$ calculated for tested numbers of degrees of freedom.
\label{Param_lambda}
}
\end{figure}

Figure~\ref{lambda} presents results for the data collected by the irradiation of the edge of the scintillator strip. The 
$\lambda$ parameter is shown as a function of $V$ degrees of freedom of the theoretical $\chi^2$. The minimum value of $\lambda$ is obtained for thirteen degrees of freedom and is c.a. 8. Large values of $\lambda (>3)$ indicate that the data are not exactly normally distributed. The $\lambda$ parameter was evaluated for the theoretical $\chi^2$ with 13 degrees of freedom for all positions. The $\lambda$ values vary between 7 and 12. From this analysis one infers that the data in each set may be approximated with 13 independent and normally distributed variables. We will follow this assumption and investigate a simple hit-position reconstruction method based on a MVN distribution of signals. However, in the future different methods not necessarily fulfilling the normality assumption should be considered, and compared with the presented one. In the following the information about the means and covariance matrices, for all data sets, and the estimated $V$ of $\chi^2$ statistics will be used to calculate the significance of assignments ($p$-values).

\subsection{An example of the hit-position reconstruction}

\begin{figure}[h!]
	\centerline{\includegraphics[width=0.7\textwidth]{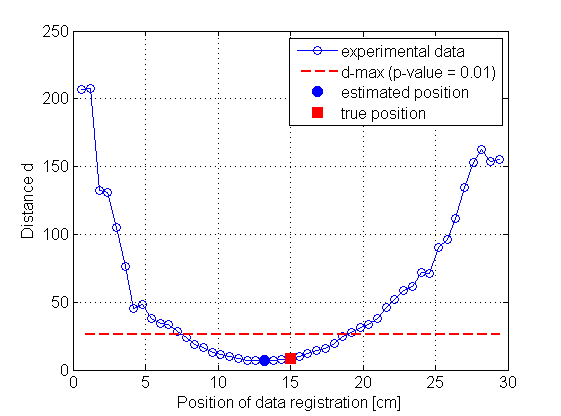}}
\caption{
Example of position reconstruction for a gamma quantum hitting in the center of the scintillator.
\label{Reco_examp}
}
\end{figure}

In the previous sub-sections it was shown that the collected data samples approximately fulfill the assumption of normality so it is worth trying to apply the method for the hit-position reconstruction introduced in Sec. 2. Reconstruction is equivalent to the qualification of the signal to one of the predefined data sets established for the various positions along the scintillator. Figure~\ref{Reco_examp} shows an example of the position reconstruction for the signal created by the gamma hitting at known position, referred to as “true”. 

The distances $d$ to all data sets were calculated according to Eq.~\ref{klasyf}, and are marked in Fig.~\ref{Reco_examp} as circles. The hit-position is defined as the one for which distance $d$ acquires a minimum (full blue circle). In this example the reconstruction procedure yielded a hit-position different by 1.8~cm from the true position marked with red square. Knowing that the distance 
$d$ is derived from $\chi^2$ distribution with 13 degrees of freedom, the $p$-value of the assignment of the signal to the best-matching data set may be evaluated. The red dotted line in Fig.~\ref{Reco_examp} indicates the maximum acceptable value of $d_{\text{max}}$=27.7 corresponding to the $p$-value threshold of 0.01. The statistical significance of the assignment of a given measurement to the best-matching data set allows for the filtering of the data sample. Because of the multiple interactions of gamma quanta, the analyzed signal may be produced as a random coincidence of two signals in different positions. Hence, the $p$-value, or equivalently distance $d$, analysis helps to reject such distorted signals and improve the reconstruction process.

\subsection{Spatial resolution of the hit-position reconstruction}

\begin{figure}[h!]
	\centerline{\includegraphics[width=0.7\textwidth]{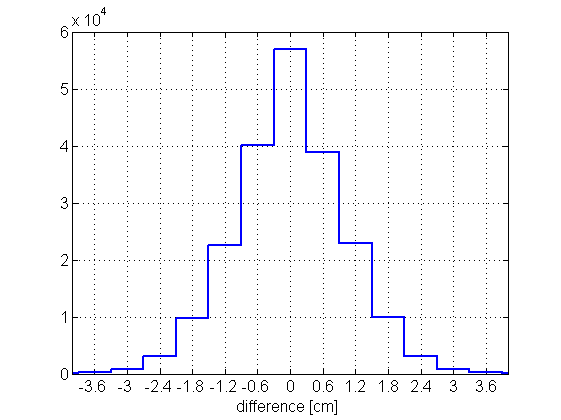}}
\caption{
Distribution of distances between true and reconstructed hit-position.
\label{Results}
}
\end{figure}

The reconstruction method was verified using signals from all data sets from all the irradiation positions. Due to the fact that each data set $S_i$ consists of about 5~000 measurements, the influence of the single measurement on parameters ($m_i$, $C_i$) of set $S_i$ is negligible. Hence, all signals in each data set $S_i$ were used to evaluate the parameters of a MVN distribution defined by ($m_i$, $C_i$). The hit position was reconstructed for each signal using the method presented in Sec.~2. Knowing the true hit-position for each measurement, differences between the true and the reconstructed positions ($r$) were evaluated (cf. Fig.~\ref{Results}).

In our method only the index of the best-matching data set is found, hence the $r$ acquires discrete values. The standard deviation ($\sigma$) of $r$ is equal to 1.05~cm. Results presented in Fig.~\ref{Results} were obtained after filtering data provided the $p$-value is larger than 0.01 and the amplitude of the signal is larger than 0.6~$A_{\text{max}}$, where 
$A_{\text{max}}$ corresponds to the amplitude at the Compton edge observed at given position of irradiation (see Fig.~\ref{Ampl_distrib}). The last criterion is used in order to reject signals with the small number of photoelectrons which spoil the resolution and anyhow are discarded in the image reconstruction in order to filter out the scattering of annihilation quanta inside the diagnosed patient~\cite{StripPETconcept,TOFPETDetector}. In comparison, the proposed method using the lowest threshold (60~mV) alone, under the same filtering conditions, gives 1.08~cm ($\sigma$) spatial resolution. In the case of using the highest threshold level (240~mV) alone the spatial resolution of 1.25~cm ($\sigma$) is obtained.

\subsection{Time resolution of the event time reconstruction}

The resolution of the time difference ($\Delta t$) between the signal arrivals to the scintillator ends may be derived directly from the previously calculated value of the spatial resolution,
\begin{equation}
	\sigma (\Delta t) = \sigma (r) \frac{2}{v_{\text{eff}}},	
\end{equation}
where $v_{\text{eff}}$ denotes the effective speed of light signal in used scintillator strip. In the recent work \cite{NIM_Mosk}, the speed of the light in the scintillator was estimated to 12.6~cm/ns. Hence, the resolution of ($\Delta t$), may be estimated to be about 167~ps ($\sigma$). This corresponds to a twice better resolution of about 83 ps ($\sigma$) for the determination of the interaction moment of the gamma quanta hit the scintillator ($t_{\text{hit}}$). The interaction moment is given by:
\begin{equation}
	t_{\text{hit}} = \frac{t_{\text{L}} + t_{\text{R}}}{2} - \frac{D}{v_{\text{eff}}},	
	\label{t_hit}
\end{equation}
where $t_{\text{L}}$ and $t_{\text{R}}$ are the arrival times to the left and right photomultipliers and $D$ is the length of whole strip. We assume for the sake of simplicity that $v_{\text{eff}}$ in Eq.~\ref{t_hit} is known exactly. Since the time difference $\Delta t = t_{\text{R}} - t_{\text{L}}$, we have
\begin{equation}
	\sigma^2 (\Delta t) = \sigma^2(t_{\text{R}}) + \sigma^2(t_{\text{L}})	
\end{equation}
and the resolution of $t_{\text{hit}}$ based on Eq.~\ref{t_hit} may be expressed as:
\begin{equation}
	\sigma^2 (t_{\text{hit}}) = \frac{\sigma^2(t_{\text{R}}) + \sigma^2(t_{\text{L}})}{4} = \frac{\sigma^2 (\Delta t)}{4}	
\end{equation}
which implies that $\sigma(t_{\text{hit}}) = \sigma(\Delta t)/2$.

\section{Conclusions}

In this article a novel method for hit-position reconstruction in plastic scintillator detectors was introduced. It was validated for the application in a new Positron Emission Tomography based on plastic scintillator 
strips~\cite{NovelDetectorSystems,StripPETconcept,TOFPETDetector,JPET-Genewa}. The method can be applied to detectors enabling sampling of signals in the voltage domain. The distinctive feature of the proposed reconstruction technique is the determination of the Mahalanobis distances of the multivariate vector, representing the measured signal, from the vectors corresponding to the mean of the signals in data sets determined for known positions. The covariance matrix and the mean vector are calculated separately for each position. The reconstruction algorithm identifies the data set for which the Mahalanobis distance acquires a minimum and reconstructs the hit position as that where the identified data set was generated. 

	The procedure was tested using a large statistics sample of data registered by a dedicated setup in 50~ps intervals. Experimental setup provided irradiation of a chosen position in a plastic scintillator strip with 511~keV gamma quanta. Sampling in the voltage domain at four thresholds was simulated and each measurement was represented by a 15-dimensional vector holding information about the relative time values of signal's arrival to both scintillator ends. 

	Using the introduced reconstruction procedure, the spatial and time resolutions of the hit-position and event time for annihilation quanta measured with a 30~cm plastic scintillator strip with sampling of signals at two edges, was determined to be about 1~cm and 80~ps ($\sigma$), respectively. It should be noted that these resolutions can be slightly improved by decreasing smearing due the finite size of the beam (0.2~cm) and due to 0.6~cm step used for the determination of data sets. 

	The performance of the method has been validated on a single scintillator strip. However, an independent test for about 20 strips have been made and the results were stable in the sense of obtained resolution ($\sigma$). Anyhow, in order to avoid inhomogeneity of response in a full scale detector, separate calibrations of each module (scintillator, photomultimpiers and electronics channels) will be provided. Moreover, the worsening of electronic performance for a full scale detector will be limited since the readout system is based on well recognized and tested components used in the particle physics experiments e.g. 
HADES~\cite{Hades}. 

	It was shown that the measured signals may be approximated with a MVN distribution with thirteen degrees of freedom. It is worth noting that the developed statistical test is general and may be incorporated in any other investigation where confirmation of multidimensional normality is needed. Since the $\lambda$ parameter (see Eq.~\ref{lambda}) is greater than 3, the further improvements in spatial and time resolutions can be achieved by applying different reconstruction methods where the assumption about normality is not obligatory, e.g. artificial neural networks. Furthermore, the resolution can be still improved by the optimization of threshold levels, an increase of their number, and enhancing light collection efficiency by optimizing the shape of scintillators and usage of silicon photomultipliers.

\section{Acknowledgements}
We acknowledge technical and administrative support of T. Gucwa-Ry\'s, A. Heczko, M. Kajetanowicz, G. Konopka-Cupia\l, 
J. Majewski, W. Migda\l, A. Misiak, and the financial support by the Polish National Center for Development 
and Research through grant INNOTECH-K1/IN1/64/159174/NCBR/12, 
the Foundation for Polish Science through MPD programme, 
the EU and MSHE Grant No. POIG.02.03.00-161 00-013/09, 
Doctus - the Lesser Poland PhD Scholarship Fund, 
and Marian Smoluchowski Krak\'ow Research Consortium "Matter-Energy-Future". We thank A. Palladino for the proofreading of the text.


%

\begin{thebibliography}{99}

%
\bibitem{Bailey2005} D. L. Bailey, 
         NJ: Springer-Verlag (2005)

%
\bibitem{Humm2003} J. L. Humm et al., 
         Eur.  J.  Nucl.  Med.  Mol.  Imaging 30 (2003) 1574
%
\bibitem{Brownell1953} G. L. Brownell, W. H. Sweet, 
         Nucleonics 11 (1953) 40
%
\bibitem{Sweet1951} W. H. Sweet, 
         New England Journal of Medicine 245 (1951) 875
%
\bibitem{Robertson1973} J. S. Robertson et al.,
         Tomographic Imaging in Nuclear Medicine (1973) 142
%
\bibitem{Karp2008} J. S. Karp et al., 
         J.  Nucl.  Med. 49 (2008) 462
%
\bibitem{Kardmas2009} D. J. Kardmas et al., 
         J.  Nucl.  Med. 50 (2009) 1315
%
\bibitem{Townsend2004} D. W. Townsend., 
         Ann. Acad. Med. Singapore 22 (2004) 133
%
\bibitem{Moses1999} W. W. Moses, S. E. Derenzo,
        IEEE Trans. Nucl. Sci. 46 (1999) 474
%
\bibitem{Moses2003} W. W. Moses,
        IEEE Trans. Nucl. Sci. 50 (2003) 1325
%
\bibitem{Conti2009} M. Conti, 
         Phys. Med. 25 (2009) 1
%
\bibitem{Schaart2010} R. D. Schaart et al., 
        Phys.  Med.  Biol. 55 (2010) N179
%
\bibitem{MoszynskiSzczesniak2011} M. Moszynski, T. Szczesniak,
             Acta Phys. Pol. B. Proc. Supp. 4 (2011) 59
%
\bibitem{Szczesniak2010} T. Szczesniak et al.,
             IEEE Trans. Nucl. Sci. 57 (2010) 1329
%
\bibitem{AmpSampl} H. Kim et al., 
          Nucl. Instr. \& Meth. A 602 (2009) 618
%
\bibitem{AX-PET} C. Casella et al., 
          Nucl. Instr. \& Meth. A 736 (2014) 161
%
\bibitem{NovelDetectorSystems} P. Moskal et al.,
  Bio-Algorithms and Med-Systems 7 (2011) 73
%
\bibitem{StripPETconcept} P. Moskal et al., 
  Nuclear Medicine Review 15 (2012) C68
%
\bibitem{TOFPETDetector} P. Moskal et al.,
  Nuclear Medicine Review 15 (2012) C81
%
\bibitem{JPET-Genewa} P. Moskal et al.,
  Radiotheraphy and Oncology 110 (2014) S69
%
\bibitem{palka} M. Palka et al.,
  Bio-Algorithms and Med-Systems 10 (2014) 41
%
\bibitem{Mardia} K. V. Mardia,
		Biometrika. 57 (1970) 519
%
\bibitem{Lillie} H. W. Lilliefors,
 		J. Am. Stat. Assoc. 62 (1967) 399
\bibitem{Fatto} L. Fattorini,
		Statistica. 2 (1986) 209
%
\bibitem{Villa} J. Villasenor Alva, E. González Estrada,
		Comm. Stat. Theory Methods 38 (2009) 1870
%
\bibitem{qqplot} M. B. Wilk, R. Gnanadesikan,
		Biometrika 55 (1968) 1
%
\bibitem{Mulnor} R. A. Johnson et al.,
        3rd. ed. New-Jersey: Prentice Hall. (1992) 158
%
\bibitem{PCA} D. P. Berrar et al.,
 		Kluwer Academic Publishers (2002)
%
\bibitem{person} K. Pearson, 
		Philosophical Magazine Series 50 (1900) 157
%
\bibitem{Szym} K. Szymanski et al., 
		Bio-Algorithms and Med-Systems (in print); e-Print arXiv:1312.0250 		
%
\bibitem{EJ} Eljen Technology http://www.eljentechnology.com
%
\bibitem{Forsyt} G. E. Forsythe et al., 
		Englewood Cliffs, NJ: Prentice-Hall (1977)
%
\bibitem{NIM_Mosk} P. Moskal et al., 
         submitted to Nucl. Instr. \& Meth. A
%
\bibitem{Hades} J. Michel et al.,
             IEEE Trans. Nucl. Sci. 58 (2011) 1745
%
\end{thebibliography}
\end{document}